\documentclass[a4paper,11pt]{article}
\usepackage{pos}
\usepackage{slashed}
\newcommand{\beq}{\begin{equation}}
\newcommand{\eeq}{\end{equation}}
\newcommand{\Lag}{\mathscr{L}}
\newcommand{\Tr}{\,{\rm Tr}\,}
\def\U{{\rm U}}
\def\SU{{\rm SU}}
\def\SL{{\rm SL}}
\newcommand{\GeV}{\,{\rm GeV}}
\newcommand{\TeV}{\,{\rm TeV}}
\newcommand{\fig}[1]{~\ref{fig:#1}}
\newcommand{\med}[1]{\langle #1\rangle}
\newcommand{\eq}[1]{~{\rm (\ref{eq:#1})}}
\renewcommand{\Re}{{\rm Re}\,}
\renewcommand{\Im}{{\rm Im}\,}

\makeatletter
\def\hhref#1{\href{http://arxiv.org/abs/#1}{arXiv:#1}}
\usepackage{xstring}
\newcommand{\hhrefq}[1]{\IfSubStr{#1}{:}{\href{http://inspirehep.net/search?ln=en&ln=en&p=#1&of=hb&action_search=Search&sf=&so=d&rm=&rg=25&sc=0}{InSpire:#1}}{\hhref{#1}}}

\def\art{\@ifnextchar[{\eart}{\oart}}
\def\eart[#1]#2#3#4#5#6{{\rm #2}, {\em #3 \bf #4} {\rm (#6) #5} ({\em #1})}
\def\article{\@ifnextchar[{\earticle}{\oarticle}}
\def\oarticle#1#2#3#4#5#6{{\rm #1}, {\em `#6'}, {\rm #2 #3 (#5) #4}}
\def\earticle[#1]#2#3#4#5#6#7{{\rm #2}, {\em `#7'}, {\rm #3 #4 (#6) #5}  [\hhrefq{#1}]}
\def\hepart[#1]#2{{\rm #2, \sl#1}}
\def\heparticle[#1]#2#3{#2, {\em `#3'} [\hhrefq{#1}]}
\newcommand{\doi}[1]{\href{http://dx.doi.org/#1}{[link]}}

\newcommand{\hhrefqq}[1]{\IfBeginWith{#1}{10.}{\href{https://doi.org/#1}{doi:#1}}{\hhrefq{#1}}}

\title{Solving the strong CP problem}
\author{Alessandro Strumia}
\affiliation{Dipartimento di Fisica, Universit\`a di Pisa,\\ Largo Bruno Pontecorvo 3, 56127, Pisa, Italia}
\emailAdd{alessandro.strumia@unipi.it}
\abstract{I briefly review solutions to the strong CP problem based on axions, parity invariance, CP-invariance,
and present a new idea based on CP as part of a spontaneously broken flavour symmetry such as a U(1) or modular invariance.}
\FullConference{The XVIth Quark Confinement and the Hadron Spectrum Conference (QCHSC24)\\
 19-24 August, 2024\\  Cairns Convention Centre, Cairns, Queensland, Australia\\}

\tableofcontents

\begin{document}
\maketitle

\section{Introduction}
QCD is described by the Lagrangian
\beq \Lag_{\rm QCD} = \sum_q \bar q (i \slashed{D} - {M_q})q - \frac14 \Tr  G_{\mu\nu}G^{\mu\nu}  + {\theta_{\rm QCD}} 
\frac{g_3^2}{32\pi^2} \Tr\, G_{\mu\nu}\tilde{G}^{\mu\nu}\eeq
where  \( G \) is the $\SU(3)_c$ gluon field strength tensor, \( \tilde{G} \) is its dual, and $q$ are quark fields.
$ \Lag_{\rm QCD} $ contains two complex CP-violating terms. Data show they have vastly different values:
\begin{itemize}
\item A combination of phases in the quark mass matrices $M_q$ is large, giving
the CKM phase $\delta_{\rm CKM}\sim 1$.
\item The rephasing-invariant combination 
$\bar\theta=  {\theta_{\rm QCD}} +\arg\det {M_q} $ must be small,
$|\bar\theta|\lesssim 10^{-11}$,
to satisfy the experimental bound on the
 neutron electric dipole, $|d_n| \lesssim1.8~10^{-26}e\,{\rm cm}$~\cite{2001.11966}.
\end{itemize}
The strong CP issue arises because generic large phases in $M_q$ typically contribute to a large $\bar\theta$.
The possible interpretations fall in two classes:
\begin{itemize}
\item[1)] Solutions based on new physics at QCD energies:
\begin{enumerate}
\item[1a)] An extra pseudo-scalar, the axion, coupled as \( a \Tr G \tilde{G} \).
\end{enumerate}

\item[2)] New physics at much larger energies could leave a $M_q$ with a special structure 
such that $\delta_{\rm CKM}\gg |\bar\theta|$. 
This could happen if some symmetry of $\Tr G\tilde{G}$ is a symmetry of the full theory, and it is suitably broken:
\begin{enumerate}
    \item[2a)] \(\Tr G \tilde{G} \) is odd under parity P: this could be a symmetry spontaneously broken by the vacuum expectation value of 
    a  real scalar field. 
    \item[2b)] \( \Tr G \tilde{G} \) is odd under CP: this could be a symmetry spontaneously broken by the vacuum expectation values of 
     complex scalar fields. 
\end{enumerate}
\end{itemize}
A third class, based on anthropic selection, is missing, because no anthropic boundary
seems to forbid a much larger $\bar\theta$.

\begin{figure}[t]
$$\includegraphics[width=0.7\textwidth]{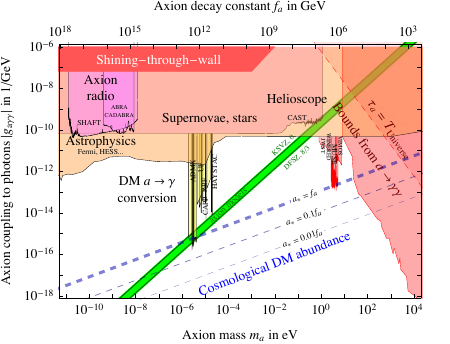}$$
\begin{center}
\caption{\em\label{fig:axion}Summary of axion searches, from~\cite{2406.01705}.
The green band in the plane (axion mass, axion coupling to photons) is favoured by axion models.
Axions produced from the misalignment mechanism match the DM abundance 
along the blue dashed lines, for different vales of the
initial axion vacuum expectation value $a_*$.
All the other shaded regions are exclusion bounds.
}
\end{center}
\end{figure}

\section{The axion solution}
The axion is a well known topic, summarised in reviews~\cite{2003.01100,2406.01705}.
Briefly, the axion $a$ would be the pseudo-Goldstone of a {\em global anomalous} U(1)  symmetry
spontaneously broken at a scale $f_a \gtrsim 10^9\GeV$, giving rise to the effective
low-energy action at the QCD scale
\beq \Lag_{\rm axion}^{\rm eff} = \Lag_{\rm QCD}+ \frac{(\partial_\mu a)^2}{2} -
\frac{a}{f_a}
\bigg[\frac{\alpha_{3}}{8\pi} \Tr G_{\mu\nu} \tilde G^{\mu\nu}+
c_\gamma \frac{\alpha}{8\pi} F_{\mu\nu} \tilde F^{\mu\nu}\bigg]+\cdots\eeq
The axion dynamically adjusts $\bar\theta=0$ acquiring a mass
$m_a \sim \Lambda_{\rm QCD}^2/f_a \approx 5.7 \, \mu{\rm eV} ~( {10^{12}\GeV}/{f_a})$.
The axion could be Dark Matter with cosmological abundance arising from the `misalignement' mechanism as
$\Omega_a \approx 0.15  ({f_a}/{10^{12}\GeV})^{7/6}({a_*}/{f_a})^2$,
plus a possible extra significant contribution from topological defects. 
The most sensitive experimental searches for axions assume that the axion is all Dark Matter and has
a coupling to photons, $g_{a\gamma\gamma} \equiv { c_\gamma \alpha}/{2\pi f_a}$.
Such experiments are reaching the region where a signal is expected, as illustrated in fig.\fig{axion}.
A future discovery would allow to measure the axion mass $m_a$ and the coupling to photons.
In a fundamental theory $g_{a\gamma\gamma}/m_a$ tells the ratio of electric  $q_f$ vs color $T_f$ charges of  fermions $f$
that contribute to the
anomalous axion couplings to photons and gluons:
\beq  \frac{g_{a\gamma\gamma}}{m_a} = \frac{2.1 \alpha}{2\pi f_\pi m_\pi} \bigg(\frac{E}{N}
-1.92\bigg),\qquad
\frac{E}{N}= \frac{{\sum_f q_f^{\rm PQ} q_f^2}}{{\sum_f q^{\rm PQ}_f T_f^2}}.
\eeq
The numerical factors are ratios of light quark masses.
SM fermions contribute as $E/N=8/3$ and the same $E/N$ arises adding heavy quarks in full multiplets of SU(5)-like unification theories.
However, axion theories suffer of the {\em quality problem}: global symmetries are expected to be accidental
and broken by gravity, that could manifest as dimension $d$ operators with coefficients
$1/M_{\rm Pl}^{d-4}$ suppressed by the Planck mass.
PQ-breaking operators ruin the axion solution if they provide a larger contribution to the axion
potential than QCD,
$f_a^4 (f_a/M_{\rm Pl})^{d-4}\gtrsim 10^{-10}\Lambda_{\rm QCD}^4$.
To avoid this, operators up to dimension $\gtrsim9$ must be much smaller than Planck-suppressed.
Such a high-quality accidental U(1) global symmetry only arises in special models.

\section{Solutions based on parity}
This approach needs significant extensions of the Standard Model, because 
weak interactions significantly violate parity~\cite{P}.
Typical models extend weak $\SU(2)_L$  into $\SU(2)_L\otimes\SU(2)_R$
broken by Higgs doublets $H_L$, $H_R$ with vacuum expectation values $v_L$, $v_R$.
Imposing a permutation symmetry between left and right,
and adding extra heavy quarks $Q$ (for simplicity singlets under the $\SU(2)_{L,R}$),
the extended quark mass matrices take the form~\cite{P}
\beq
 M_q = \bordermatrix{ &q_R & Q_R \cr
q_L &  0 & {y_L}  v_L \cr
Q_L & {y_R} v_R &M}.\eeq
The left $\leftrightarrow$ right symmetry implies $y_R = y_L^\dagger$, such that
this matrix has real determinant,
despite that the Yukawa couplings $y_{L}$ are generically complex.
However, a real $\det M_q$ also needs that the Higgs vacuum expectation values
$v_L,v_R$ are real.
This is not generically the case, as the Higgs potential
$V(H_L, H_R)$ can contain complex CP-violating couplings.
To avoid this, the potential $V$ can be restricted imposing supersymmetry.
Then, loop corrections to $\bar\theta$ are small enough if $v_R \gtrsim 10\TeV$,
as the neutron electric dipole $d_n \sim e g^2 m_q/(4\pi v_R)^2$ arises at one loop,
and a correction to $\bar\theta \sim (y_t/4\pi)^4 v_L/v_R$ at two loops.

\smallskip

Variants of this basic idea have been proposed,
such as a mirror copy of the SM with gauge group $\SU(2)_L\otimes\SU(2)'_L$.

\section{Solutions based on CP}
Imposing CP needs milder SM extensions and allows to address a more general interesting issue: 
why the SM action contains complex parameters?
Typical models assume CP invariance (so, all parameters are real)
spontaneously broken by `complex' scalar singlets $z$.
Nelson and Barr showed that special mass matrices involving heavy quarks $Q$ and having special 0 entries
have real determinant
\beq \label{eq:NB}
M_q\sim\bordermatrix{ & q_R &  Q_R \cr
q_L & y_q v & {0} \cr
Q_R^c & y  \med{{z}} +y'  \med{z}^*  & M}\
\eeq
such that $\bar\theta=0$ at tree level, while the CKM phase can be large.
Bento et al.~\cite{Bento:1991ez} realised this idea in concrete models, based on the interaction
\beq
 y_q H q_L q_R+ (y z+ y' z^*) Q_R^c q_R + M Q_R Q_R^c - V(H,z).\eeq
The 0 entry is realised by forbidding 
a $H q_L Q_R$ interaction by imposing a $\mathbb{Z}_2$ that flips $Q_R, Q_R^c, z$.
However, this symmetry does not forbid higher-dimensional operators such as
$z H q_L Q_R /M_{\rm Pl}$. Assuming that such operators are present but suppressed by the Planck scale,
their effect is small enough provided that CP is broken at low enough energy,
$\med{z} \lesssim \bar\theta M_{\rm Pl}\sim 10^8 \GeV$.
Loop corrections to $\bar\theta \sim \lambda_{Hz} y^2/(4\pi)^2$ can be too large, 
and can be suppressed by either assuming small couplings 
(such as a small mixed scalar quartic $\lambda_{Hz}$)
or supersymmetry.

\subsection{Solving the strong CP problem using CP and a U(1) symmetry}
I present a new solution, that looks to me plausible.
Let's assume that CP is (part of) a spontaneously broken flavour symmetry.
The plausible realization arises employing, as flavour symmetry,  a modular symmetry~\cite{2305.08908}.
For ease of presentation, I introduce the idea assuming a conventional
U(1) flavour symmetry~\cite{2406.01689} broken by scalars $z_a$ with given charges $k_{z_a}$. 
These charges are unusually denoted as $k_{z_a}$ because they will later become modular weights.
The $\med{z_a}$ vacuum expectation values of CP-breaking scalars are the only complex quantities in the theory,
in particular $\theta_{\rm QCD}=0$.
As explained and justified later, we need to assume that the elements $m$ of the quark mass matrices are given by
\beq \displaystyle m = \sum  \hbox{(real constants $c$)} \, \prod_a z_a ^{k_a}
\eeq
with {\em positive integer powers} $k_a\ge 0$ and {\em no entries proportional to the conjugated} $z_a^\dagger$.
The sum runs over possible multiple choices of $k_a$.
As a toy example, we consider one $z$ and $N_g=2$ generations of SM quarks, $q_{L1,2}$ and $q_{R1,2}$.
Following our assumptions, the quark mass matrix is
\beq M_q =  \bordermatrix{ & q_{R1} & q_{R2}\cr
q_{L1} & c_{11} z^{k_{11}}&c_{12}z^{k_{12}}\cr
q_{L2} & c_{21} z^{k_{21}}&c_{22}z^{k_{22}}}\eeq
where $c$ are real constants and the powers are obtained as
$k_{ij} = (k_{q_{Li}} + k_{q_{Rj}} + k_{H_q})/k_z$
by matching the U(1) charges $k$ of the various fields.
Then, the determinant of the matrix simplifies into
\beq
\det \label{eq:detMtoy}
M_q=c_{11} c_{22} \, z^{k_{11}+k_{22}}-c_{12} c_{21} \, z^{k_{12}+k_{21}} = (c_{11} c_{22}- c_{12} c_{21}) z^k.\eeq
This shows that $\det M_q$ can be real for any values of the real constants $c$ provided that the
`total U(1) charge' $k$ of the fields involved in the determinant vanishes:
\beq\label{eq:ksimpl} k = \sum_{i=1}^{N_g=2} (k_{q_{Li}} + k_{q_{Ri}} + k_{H_q})=0.\eeq
As a consequence $\det M_q$ is real and $\bar \theta=0$. 
This is not yet a solution to the strong CP problem, because the CKM phase trivially vanishes.
The same would happen with $N_g=3$ generations:
in general, CP remains unbroken when a U(1) symmetry is broken by one scalar $z$ only,
as its phase is unphysical. 
A U(1) transformation allows to set $\med{z}$ to a real value, without loss of generality.
CP is instead broken if the flavour U(1) is broken by multiple scalars $\med{z_a}$ with different phases:
their relative phases are physical.

The key property of the toy model in eq.\eq{detMtoy} remains non-trivially true in general:
the determinant is a special object that still satisfies the  identity 
\beq \det M_q  \propto \lambda^k\qquad \hbox{when}\qquad z_a\to \lambda^{k_a} z_a\eeq
for a generic number of scalars $z_a$ and for a generic number of quarks ($N_g$ generations, plus possibly heavy quarks).
So, similarly to eq.\eq{ksimpl}, $\det M_q$ is real if the total charge vanishes:
\beq \label{eq:k}
k=\sum_i (2k_{Q_i} +k_{U_i} + k_{D_i}+ k_{H_u}+k_{H_d})=0.\eeq
Here $Q=(u_L, d_L)$ are the left-handed SM quark doublets; $U,D$ are right-handed quarks,
and $H_q$ are Higgs doublets that give mass to up-type and down-type quarks
(a unique Higgs doublet $H_u=H_d^\dagger$ is present in the SM, while two Higgs doublets are needed in supersymmetric
extensions).

\smallskip

This class of theories provides a solution to the strong CP problem, as they allow for a large $\delta_{\rm CKM}\sim 1$. 
As a concrete example, let's use consider the simplest model with no heavy quarks and just the SM
$N_g=3$ generations of chiral quarks.
In such a case this logic leads to a unique Yukawa matrix (up to permutations, possible degeneracies or extra vanishig entries~\cite{2305.08908,2404.08032,2406.01689}):
\beq  Y = \left(
\begin{array}{ccc}
 0 & 0 & c_{13} \\
0 & c_{22} & Y_{23} \\
c_{31} &  Y_{32} &
Y_{33}
 \end{array}
\right),\qquad \det Y = -c_{13}c_{22}c_{31}.
\eeq
This form can be for example realised assuming that U(1) is broken by two scalars $z_{+}$ and $z_{++}$ with U(1) charges $1$ and $2$.
We assume that the Higgs has no U(1) charge, and
indicate around the Yukawa matrix the negative of the U(1) charges of the quarks $Q$ and $U$ or $D$:
\beq Y = \bordermatrix{ & -1 & 0 & +1\cr
-1 &  0 & 0 & c_{13} \cr
\phantom{-}0 & 0 & c_{22} & c_{23}\, z_{+} \cr
+1 & c_{31} &  c_{32} z_+& c_{33} \, z_+^2 + c_{33}^{'}  z_{++}}.
\eeq
This notation means that a U(1)-invariant $23$ entry needs a scalar $z$ with total charge $+1$,
so the only possibility is $z_+$ times a real coefficient $c_{23}$.
The 11, 12, 21 entries of the Yukawa matrix need scalars with total negative charge and thereby
vanish because of our
(so far unjustified) assumption that powers of $z$
cannot be negative and that complex conjugated scalars $z^\dagger$ cannot appear.\footnote{Similar Yukawa matrices
motivated by the QCD $\theta$ problem have been considered in~\cite{hep-ph/9612396} and in \cite{1307.0710},
where the vanishing entries are obtained imposing supersymmetry and an
$A_4\otimes {\rm U}(1)_R\otimes\mathbb{Z}_2\otimes\mathbb{Z}_4\otimes\mathbb{Z}_4\otimes\mathbb{Z}_4\otimes\mathbb{Z}_4\otimes\mathbb{Z}_4$ symmetry suitably broken by 9 scalars.}
Two different terms are allowed in the 33 entry, as the needed total charge 2 can be realized either as $z_+^2$ or as $z_{++}$.
The 33 entry thereby acquires a non-trivial relative phase depending on $c_{33}/c'_{33}$, 
allowing for a non-vanishing and large CKM phase.
The determinant of $Y$ is real, so that $\bar\theta=0$.
Extra terms containing neutral combinations such as $z_{++}/z_+^2$ or $z_{++} z^{\dagger2}_+$ would induce $\bar\theta\neq 0$, 
but are not allowed because we assumed no negative powers and no conjugated fields.

As a minor variant (that will be relevant later) we again consider just the SM quarks and
two scalars $z_{4}$ and $z_{6}$ with U(1) charges $4$ and $6$, such that the Yukawa matrix has a similar structure:
\beq \label{eq:Y46}
Y = \bordermatrix{ & -6 & 0 & +6\cr
-6 &  0 & 0 & c_{13} \cr
\phantom{-}0 & 0 & c_{22} & c_{23}\, z_{6} \cr
+6 & c_{31} &  c_{32} z_6& c_{33} \, z_6^2 + c_{33}^{'}  z_{4}^3}.
\eeq
The general paradigm allows for many more possibilities, including justifying the
Nelson-Barr mass matrix of eq.\eq{NB} in models where additional heavy quarks are introduced.
Let us for example add $N_g=3$ right-handed extra heavy quarks $Q_R\oplus Q^c_R$. 
In view of $\SU(2)_L$ invariance, the $6\times 6$ quark mass matrix must have the form
\beq M _q =\bordermatrix{ &q_{R_{1,2,3}} & Q_{R_{1,2,3}}  \cr
q_{L_{1,2,3}}  & yv_q  & y'v_q \cr
Q^c_{R_{1,2,3}}  &\mu  & M}
\eeq
where the mass terms $M$ and $\mu$ can be much larger than the weak scale $v_q$.
The Nelson-Barr ansatz corresponds to assuming that $y'=0$, that  $y$ and $M$ are real, while  $\mu$ is complex.
In our context these assumptions can be justified by assuming the following U(1) charges:
\beq M_q  =\bordermatrix{ & 0 & -1 \cr
0 & yv  & 0 \cr
1 & c z  & M}.
\eeq
More general models can have complex $M, y', y$ and a light field content anomalous under the U(1) symmetry.
In the full theory $\bar\theta=0$ as the full mass matrix has a real determinant.
The same physics $\bar\theta=0$ arises in the effective field theory for light quarks only as follows.
The light quark mass matrix $M_{\rm light} $ becomes a complicated function of $z_a$
with a complex  $\det M_{\rm light}$. Its phase gets cancelled by the
anomalous contribution to gauge kinetic function $f_{\rm EFT} = f_{\rm UV} - \ln \det M_{\rm heavy}/8\pi^2$
that arises when integrating out the heavy quarks.
The two contributions together reconstruct the real determinant of the full mass matrix.
This anomaly cancellation mechanism is well known in string models with a QFT field content anomalous under
some stringy symmetries.

\smallskip

The proposed idea is so far based on assumptions.
To claim that this idea solves the strong CP problem
we need to implement and  justify the assumptions in plausible Quantum Field Theories.
Our assumption that Yukawa couplings do not depend on conjugated fields
$z^\dagger_a$ can be justified by assuming {\em supersymmetry}, 
possibly broken at energies much higher than those explored by colliders (for example, at the unification scale).
Indeed a theory invariant under (for simplicity, global) supersymmetry is described by:
\begin{itemize}
\item A holomorphic super-potential $W(\Phi)$ that depends on chiral super-fields $\Phi$, but not on their conjugates $\Phi^\dagger$:
\beq W =Y^u_{ij}(z_a) \, U_i   Q_j\, H_u  +  Y^d_{ij}(z_a)  \, D_i Q_j\, H_d + \cdots\eeq

\item Kinetic terms are described by the `Kahler' function $K(\Phi,\Phi^\dagger)$,
which is a generic real function of super-fields $\Phi$.
It does  not contribute to $\bar\theta$~\cite{CPK}.

\item The kinetic function $f$ for gauge bosons. It must be real because we assume CP.
The minimal form for gluons is $f=1/g_3^2 - i \theta_{\rm QCD}/8\pi^2$ with $\theta_{\rm QCD}=0$.
\end{itemize}
In order to avoid Goldstones and gravitational breaking of the U(1) symmetry we assume (unlike in the axion solution) that the
U(1) flavour symmetry is {\em local}.
Then, all anomalies must vanish, in particular the mixed $\U(1)\cdot \SU(3)_c^2$ anomaly with QCD
given in the Minimal Supersymmetric Standard Model by
\beq \label{eq:AU1}
A= \sum_i (2k_{Q_i}+k_{U_i}+k_{D_i})=0 .\eeq
The anomaly cancellation condition coincides with solving the QCD $\bar\theta=0$ problem by imposing eq.\eq{k}
provided that $k_{H_u}+k_{H_d}=0$, meaning that the Higgs doublets $H_{u,d}$ do not break the U(1) flavour symmetry.
In summary, $\bar\theta=0$ can be understood if CP is an anomaly-free flavour symmetry not broken by the SM Higgs doublet(s).

\medskip

Finally, implementing the general idea within full renormalizable theories needs two additional steps:
\begin{itemize}
\item[1)] A scalar potential $V(z)$ minimised by $\med{z_a}$ with different relative phases, such that CP is broken.
\item[2)] Extra  fields that mediate effective operators in a special way
such that only non-negative powers $k_a\ge 0$ appear.
\end{itemize}
Realising such conditions in supersymmetric theories with a U(1) symmetry seems possible but not nice.

\subsection{Solving the strong CP problem using CP and modular invariance}
We next show that such conditions are automatically satisfied if the same general idea is realised
using {\em modular invariance} as the CP-breaking symmetry, rather than a simpler U(1).
While modular invariance sounds to physicists as a scary topic,
it's not more dangerous than going to a physics conference in Australia 
on the wrong side of the road while looking at crocodiles, stonefish, stinging bushes.
Modular invariance is just something similar to a U(1) automatically broken in a predictive $k\ge 0$ 
way by 2 scalars with different phases. 
It automatically provides what is needed to solve the CP problem.

\smallskip

\begin{figure}[t]
$$~~\includegraphics[width=0.95\textwidth]{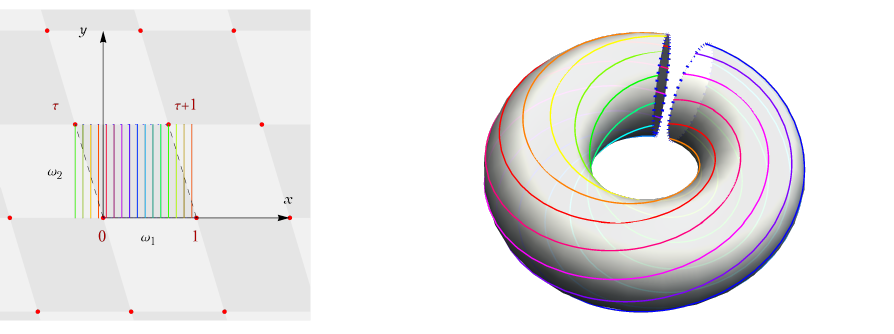}$$
\begin{center}
\caption{\em\label{fig:twist}
Basic modular transformations that leave invariant a 2-dimensional flat torus.
}
\end{center}
\end{figure}

Most authors working on modular invariance present it as the mathematics of complex functions with double periodicity.
I here present the physics motivation, emphasising that
modular invariance can be the effective field-theory manifestation
of how CP-violation plausibly arises in string theories.
This direct link makes the current approach more plausible than other ad-hoc attempts.

Briefly, super-strings theories are real  in $9+1$ dimensions.
Chiral families of fermions with CP-violating  couplings in $3+1$ dimensions 
can arise after compactifying on spaces with a property called `complex structure'.
In this context, CP can be a geometric symmetry spontaneously broken by the shape of the compactification.
Much attention has been dedicated to compactifications on Ricci-flat spaces that preserve $N=1$ supersymmetry.
The simplest example is orbi-folded flat tori in 6 dimensions.
For our purposes, it's enough to consider a 2 dimensional torus, such that the needed mathematics gets intuitive.
A 2-dimensional flat torus is simply built by folding a 2 dimensional plane $x,y$, merging
opposite edges to form two circles.
Mathematically, this operation is described by
defining a `complex structure' $z=x+i y$ and identifying points under a double periodicity
\beq z = z+n_1\omega_1 + n_2 \omega_2\eeq
for given values of $\omega_1$ and $\omega_2$ and integer $n_{1,2}$.
The absolute values $|\omega_1|$ and $|\omega_2|$ tell the periodicities along the two circles.
If $\tau \equiv \omega_1/\omega_2$ is purely imaginary, the two directions correspond to orthogonal vectors $\vec\omega_{1,2}=(\Re\omega_{1,2},\Im\omega_{1,2})$.
The resulting flat torus is well known to those who played Pac-Man on a rectangular computer screen:
by moving horizontally one exits on the right side and re-enters on the left side at the same height.
A real part of $\tau$ corresponds to a `twist': the $x,y$ space is folded in such a way
one re-enters at a different height, as illustrated in fig.\fig{twist}.\footnote{The torus on the right side can be misleading,
as the surface of the torus  in 3 dimensions is not flat: a flat 2-dimensional torus can be embedded in 4 dimensions.}
If the theory in higher 5+1 dimensions is just a real free fermion, 
the twisting affects the boundary conditions and thereby the spectrum of KK modes.
The twisting appears as CP-violation in the low-energy effective theory.

\begin{figure}[t]
$$\includegraphics[width=0.7\textwidth]{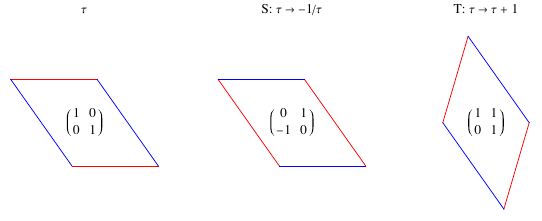}$$
\begin{center}
\caption{\em\label{fig:ST}
A 2-dimensional flat space $x,y$ is decomposed in a lattice with periodicities given by two vectors $\omega_1$ and $\omega_2$.
}
\end{center}
\end{figure}

Modular invariance is a simple consequence of this geometry:
the same lattice in fig.\fig{twist} is described by multiple values of $\omega_{1,2}$. 
The two basic transformations that leave the compactification space invariant are
illustrated in fig.\fig{ST}:
S) flipping $\omega_1\leftrightarrow -\omega_2$ ($\tau\to -1/\tau$);
T) shifting $\omega_1\to \omega_1+\omega_2$ ($\tau\to \tau+1$).
These two generators lead to the most general modular transformation 
\beq
\left(\begin{array}{cc} \omega_1 \cr \omega_2\end{array}\right)
\to
\left(\begin{array}{cc} a&b \cr c & d\end{array}\right)
\left(\begin{array}{cc} \omega_1 \cr \omega_2\end{array}\right)
\eeq
where $a,b,c,d$ are {integers} with $ad-bc=1$.
The modular $\SL(2,\mathbb{Z})$ is simply a discrete sub-group of global reparametrizations in the extra dimensions.
The higher dimensional theory, compactified on a  flat torus,
is described by a 3+1 dimensional effective field 
theory containing a field $\tau = \omega_1/\omega_2$ that parameterizes
the higher dimensional shape, with modular-invariant action.
We are interested in theories that also contain matter super-fields $\Phi$. 
They transform under modular invariance as 
\beq \Phi \to (c\tau+d)^{-k_\Phi}  \Phi\qquad
\hbox{with `weight' $k_\Phi$}.\eeq
The multiplicative factor is complex: its phase acts as a U(1) rotation
showing that the weight $k$ generalises the U(1) charge.
The U(1) is spontaneously broken because the compactification space must have some shape,
parameterised by $\tau$.
The minimal globally supersymmetric  action for $\tau$  and chiral fields $\Phi$ (quarks and Higgses)
and vectors $V$ (gluons) is
\begin{eqnarray}\label{eq:Kcan}
K &=&  -h^2 \ln(-i\tau+i\tau^\dagger) + \sum_{\Phi} \frac{\Phi^\dagger {e^{2 V}} \Phi}{(-i\tau + i \tau^\dagger)^{k_\Phi}},\\
W&=&Y^u_{ij}(\tau) U_i Q_{j} \, H_{u} +Y^d_{ij}(\tau)   D_i Q_j\, H_{d}. 
\end{eqnarray}
The parameter $h$ is analogous to the axion decay constant.
The action is modular invariant if the Yukawa couplings $Y(\tau)$ are special functions of $\tau$
that transform as
\beq Y_{ij}^q(\tau)\to (c\tau+d)^{k^q_{ij}} Y_{ij}^q(\tau)\qquad \hbox{with}\qquad
k^q_{ij}=k_{q_{Ri}}+k_{q_{Lj}} + k_{H_q}.\eeq
The needed modular weights sum like U(1) charges.

Next, some mathematics is useful.
This framework is predictive because
the only modular functions of $\tau$ with  weight $k$ and no
singularity (known as `{\it forms}') are
the Eisenstein series $E_k$, that transform nicely thanks to the lattice summation:
\beq E_{k}(\tau) \equiv \frac{1}{2\zeta(k)} \sum_{(m,n)\neq(0,0)}\frac{1}{(m + n\tau)^{k}}.\eeq
The normalization factor is just a convention.
The $E_k$ are uniquely defined for integer $k$.
The summation is divergent for $k\le 2$ (this excludes $k<0$) and vanishes if $k$ is odd.
The first two non-trivial modular forms, $E_4$ and $E_6$, have different phases.
The modular forms with $k>6$ happen to be linear combinations of $E_4$ and $E_6$, 
such as $E_8 \propto E_4^2$ and $E_{10}\propto E_4 E_6$.
So modular invariance behaves similarly to the example of eq.\eq{Y46}:
a U(1) symmetry broken by two scalars with charges 4 and 6 and a relative phase.
The modulus $\tau$ transforms as
$\tau \to- \tau^\dagger$ under CP,
so a vacuum expectation value with a non-vanishing real part breaks CP.
(One could redefine $T=i\tau$ such that the imaginary part of $T$ breaks CP).

\medskip

In summary, the strong CP problem can be solved assuming that:
\begin{itemize}
\item CP is a symmetry broken only by the modulus  ${\rm Re}\,\tau$.

\item Supersymmetry exists and is broken, possibly much above the weak scale,
by some mechanism such as gauge mediation that does not introduce extra sources
of CP violation.

\item The MSSM Higgs doublets have modular weights $k_{H_u}+k_{H_d}=0$
such that they do not break modular invariance.
\end{itemize}
Then everything follows. Summarising what was already discussed:
\begin{itemize}
\item  Yukawa couplings must be sums of real coefficients $c$ times modular forms with weight $k$,
\beq Y^q_{ij}(\tau) = \sum c^q_{ij} \, F_{k^q_{ij}}(\tau).\eeq

\item Modular invariance cannot have anomalies, since it's
just a remnant of general covariance in higher dimensions.
In particular the QCD modular anomaly must vanish. 
The anomaly coefficient is given by the same expression as the U(1) anomaly of eq.\eq{AU1}:
\beq A =\sum_{i=1}^3 (2 k_{Q_i}+ k_{U_{i}}+k_{D_{i}}) =0 \eeq
in a theory with SM quarks only.

\item Then $\det M_q$ is a modular form with weight $A=0$, so it's a real constant, so
the QCD $\bar\theta$ problem is solved as
$$ \arg \det M_u M_d=0,\qquad \theta_{\rm QCD}= 0.$$
\item The CKM phase $\delta_{\rm CKM}\propto {\rm Im}\,\det [Y_u^\dagger Y_u, Y_d^\dagger Y_d] $
has no special modular properties, and it acquires order unity values $\delta_{\rm CKM}\sim 1$ since
$E_4^3$ and $ E_6^2$ have different phases.

\item Quark kinetic matrices $Z_q$ coming from the K\"ahler function
can be made canonical $Z_q={\rm diag}(1,1,1)$ via a linear transformation that affects  quark masses and mixing while leaving $\bar\theta$ invariant. 
The minimal form of eq.\eq{Kcan} for the K\"ahler function leads, in the basis where quark kinetic terms are canonical, 
to
\beq Y^q_{ij}|_{\rm can} = \sum c_{ij}^q (2{\rm Im}\,\tau)^{k^q_{ij}/2} F_{k^q_{ij}}(\tau).\eeq

\end{itemize}
The final step is showing that this framework allows to reproduce all observed quark masses and mixings,
including the CKM phase.
The simplest model that achieves this assumes the SM light quarks only, with
modular weights $k_Q = k_{U} = k_{D}=\{-6,0,+6\}$ for the three generations.
The resulting Yukawa coupling matrix has the form of eq.\eq{Y46}:
\beq Y_{q}|_{\rm can} = 
\bordermatrix{ & q_{L1} & q_{L2} & q_{L3}\cr 
q_{R1} & 0 & 0 & c^q_{13} \cr 
q_{R2} & 0 &c^q_{22} & c^q_{23} (2\Im\tau)^{3} E_{6}(\tau)\cr 
q_{R3} & c^q_{31} & c^q_{32}  (2\Im\tau)^{3} E_{6}(\tau) &(2\Im\tau)^{6} \left[c^{q}_{33} E_4^3(\tau) 
+ {c'}^{q}_{33}  E_6^2(\tau)\right]}.\eeq
The determinant is explicitly real.
As an additional bonus, the hierarchies in quark masses and mixings can be
reproduced for comparable values of the $c^q_{ij}$ coefficients~\cite{2305.08908},
as the modular symmetry automatically leads to a Yukawa matrix 
similar to the ones obtained by Froggatt-Nielsen by introducing a U(1) flavour symmetry.
The quark mass hierarchy comes thanks to the factor 6 demanded by the modular symmetry:
e.g.\ $(2\Im\tau)^6=64$ for $\tau \sim i $.
Charged leptons and neutrinos too can be fitted in this context, assuming the same
modular weights as quarks,  $k_L=k_{E}=k_Q$,
where $L=(\nu_L, e_L)$ and $E=e_R$. 

As a final extension, this solution to the strong CP problem can be generalised to
more complicated string compactifications where the modular group
gets partially broken by the presence of branes and orbifold fixed points,
such that SM fermions have low modular weights $k=\{0,\pm 1\}$.
In mathematical language, the effective field theory retains invariance under a principal congruence
sub-group $\Gamma(N)$ of the modular group $\SL(2,\mathbb{Z})$ at higher level $N>1$.
The three generations can become components of
geometrically-broken non-abelian flavour groups 
$\SL(2,\mathbb{Z})/\Gamma(N) \sim \{S_3, A_4,S_4, A_5,\ldots\}$, 
which are automatically embedded in modular invariance.
In the lepton sector, this leads to theories of large neutrino mixing angles
that can be predictive assuming a minimal Kahler function~\cite{1706.08749}.
In the quark sector, this complicates getting small mixing angles.

Since supersymmetry plays a key role in preventing complex conjugated fields,
the scale of supersymmetry breaking must be mildly below the scale $h$ at which the modular symmetry is broken.
String compactifications lead to $h\approx M_{\rm Pl}$.
We here assumed $h \ll M_{\rm Pl}$ to avoid writing more complicated super-gravity actions.
The mechanism can be extended to super-gravity~\cite{2305.08908},
but I would not be able to present it in a simple way.

\section{Conclusions}
In conclusion, the old axion solution to the strong CP problem appears plausible
(up to the quality problem) and experimental tests are coming.
Alternative solutions based on special forms of the quark mass matrix
and no new physics at the QCD scale
seemed possible but less plausible because ad-hoc complications were needed for
their theoretical implementation.
Modular invariance seems to provide a plausible implementation.
How can this scenario be tested?
In an allowed region of parameter space  the $\tau$ modulus can be so light that it
gives flavour effects~\cite{2411.08101}.
However, the $\tau$ mass can more plausibly be just mildly below the Planck scale.
Possible cosmological signals deserve to be explored.

\section*{Acknowledgments}
I thank Arsenii Titov for comments, and the QCHSC24 
organisers for the invitation to a very interesting conference about real observable physics.

\end{document}